\documentclass[aps,twocolumn,pra,superscriptaddress]{revtex4-1}
\usepackage{graphicx}
\usepackage{amsthm}
\usepackage{amsmath}
\usepackage{amssymb}
\usepackage{times}
\usepackage{epstopdf}
\usepackage[english]{babel}
\usepackage{natbib}
\usepackage{hyperref}
\hypersetup{colorlinks=true, citecolor=blue, linkcolor=blue}
\usepackage{cleveref}
\usepackage{tablefootnote}
\usepackage{braket}
\usepackage{verbatim}
\usepackage{subfigure}
\usepackage{chngcntr}

\usepackage[pass]{geometry}

\usepackage[normalem]{ulem}
\usepackage{mathtools}

\date{November 2022}

\begin{document}

\title{Emergent Order in Classical Data Representations on Ising Spin Models}

\author{Jorja J. Kirk}
\author{Matthew D. Jackson}
\author{Philip Intallura}
\affiliation{HSBC Holdings Plc., 8 Canada Square, London, UK}
\author{Daniel J.M. King}
\affiliation{HSBC Holdings Plc., 8 Canada Square, London, UK}
\affiliation{Department of Materials, Imperial College London, Exhibition Road, South Kensington, SW7 1SZ, UK}
\author{Mekena Metcalf}
\affiliation{HSBC Holdings Plc., One Embarcadero Ctr. San Francisco, CA}
\email{mekena.metcalf@us.hsbc.com}

\begin{abstract}
Encoding classical data on quantum spin Hamiltonians yields ordered spin ground states which are used to discriminate data types for binary classification. The Ising Hamiltonian is a typical spin model to encode classical data onto qubits, known as the ZZ feature map. We assess the ground states of the Ising Hamiltonian encoded with three separate data sets containing two classes of data. A new methodology is proposed to predict a certain data class using the ground state of the encoded Ising Hamiltonian. Ground state observables are obtained through quantum simulation on a quantum computer, and the expectation values are used to construct a classical probability distribution on the state space. Our approach is a low dimensional representation of the exponentially large feature space. The antiferromagnetic ground state is the stable ground state for the one-dimensional chain lattice and the 2D square lattice. Frustration induces unique ordered states on the triangle lattice encoded with data, hinting at the possibility for an underlying phase diagram for the model. We examine order stability with data scaling and data noise. 
\end{abstract}

\maketitle

\section{Introduction}
Mapping classical data onto quantum spin models underpins many quantum machine learning frameworks, specifically, where one wants to determine optimum generators to express data relationships~\cite{Schuld2021}. This is a difficult task, since the symmetries in quantum data don't translate over to classical data. Encoding data on spin lattices inherently represents unstructured classical data on a quantum system, therefore, can we understand more about classical data by using our quantum simulation toolbox? Recent work suggest Hamiltonian frequencies can elucidate whether a quantum machine learning model will generalize in the presence of noisy data \cite{Peters2022}. Inspired by the role the Hamiltonian spectra plays in the effectiveness of quantum machine learning models, we decided to analyze the spectra of classical data mapped to the standard Ising model known as the ZZ feature map. We find the ground state Eigen vector reveals an interpretable representation for classical data mapped to quantum spin lattices that can discriminate against classes, analyzing the role of noisy data, and the role of data scaling. 

Merging quantum information and machine learning provides a platform to explore new models and paradigms to classify classical data  ~\cite{SchuldPetruccioneBook}. Precise simulation of Ising models is classically hard, therefore, quantum advantage is provable using the spin lattice as a feature map \cite{Havlicek2019}. Machine learning models are scrutinized by scaling, accuracy, and interpretability. Quantum computing typically is used to address scaling in linear models while accuracy, when theoretical error bounds are absent, is left to empirical investigations with classical data sets. Model interpretability, excluding linear solvers, is less addressed in the context of quantum machine learning, since quantum neural networks and other variational routines are heuristic in nature.  Model interpretability is critical when making high stakes decisions like medical diagnosis or identifying criminal activity~\cite{Gilpin2018,Rudin2019}. Eigen states of feature maps are not heuristic and provide information about classical data in quantum representations. Projecting classical data on the feature map Eigen space is an interpretable technique to gain insights about the data and potentially discriminate data. 

We investigate three different data sets used to test classifier performance; these include the UC Irvine breast cancer data set \cite{UCIBCData}, credit card fraud detection data set~\cite{KaggleCCT}, and e-commerce data set~\cite{EcommData}. Data is encoded on 1D and 2D Ising spin lattices, and we evaluate the ground state of these data Hamiltonians using exact diagonalization. We find the ground state solution yields an antiferromagnetic phase with two non-degenerate antiferromagnetic states. We build a statistical distribution by evaluating the model ground states over a subset of data. We compare the distributions of the two classes using total variation distance and find model success is correlated to the large distribution differences. Our results imply antiferromagnetic order is useful for data class discrimination. We address order stability with noisy data, and find it is robust to small amounts of noise. We also find correlation is influenced by data scaling which explains the need for the scaling parameter used in feature encodings. 

This paper is arranged as follows: in Section 2 we describe feature encoding on the Ising model. Section 3 includes our proposed methodology to classify data using ground state evaluation with classical distribution sampling. Section 4 contains numerical results of three data sets encoded on one-dimensional and two-dimensional spin lattices. Model robustness with scaling and noise is quantified in Section 5.


\section{Data Encoding Feature Maps}
\label{section:encode}
There exist three paradigms to encode classical data vector, $\vec{x}$, to quantum computers: angle encoding, basis encoding and amplitude encoding~\cite{Schuld2019}. Amplitude encoding compactly maps a data set onto a wave-function with $\mathcal{O}(log(MN))$ qubits where $M$ is the number of features and $N$ is the number a vector. In the worst case when the data lacks structure, amplitude encoding is linear in the data vector dimension $\mathcal{O}(MN)$. Angle encoding, where each qubit represents a datum feature, is a near term amenable encoding approach. Mapping classical data onto a quantum spin model is an encoding approach connecting quantum simulation to quantum machine learning. Hamiltonian encoding forces classical data to become quantum in nature through re-representation as a disordered spin lattice. It is difficult to surmise any advantage apart from scaling without intuition for the nature behind this re-representation. In some cases, finding the optimal representation is left to combinatorial optimization problems and clever genetic algorithms~\cite{AL2021, Incudini2022, Torabian2022}, however, this does not provide insight into the models ability to capture non-linearity and discriminate data.

A foundational quantum machine learning paper, Havlicek et. al. Ref~\cite{Havlicek2019}, developed a data encoding strategy to perform quantum classification algorithms. Data non-linearity is enveloped by spin interactions on a lattice and non-linearity expressed as entanglement is beyond the reach of classical kernel routines. Consider a data vector, $\vec{x}$, with $N$ features mapped to $N$ qubits. A feature map is constructed with interaction degree $S \subseteq [N]$ as $V = U(\vec{x})\mathcal{H}^{\otimes N}U(\vec{x})\mathcal{H}^{\otimes N}$ where $\mathcal{H}$ is the Hadamard gate and 
\begin{align}
\label{Eq:Hamiltonian}
U(\vec{x}) &= e^{iH}\\
H &= \sum_{S\subseteq [N]} \phi_S(\vec{x}) \prod_{i \in S} \hat{Z}_i.
\end{align}
The function $\phi_S(\vec{x})$ encodes the data onto the coefficients of Pauli operations. When $S\leq 2$ the coefficients are 
\begin{align}
\phi_i(\vec{x}) &= x_i\\
\phi_{i,j}(\vec{x}) &= \left(\pi - x_i\right)\left(\pi - x_j\right).
\end{align}
This encoding strategy is commonly known as the ZZ feature map whose pairwise interactions yield an Ising type Hamiltonian. In many real world cases, classical data has no structure or symmetry, thus, the resulting Ising Hamiltonian represents a disordered spin lattice. What information can we gain about the classical data by investigating the disordered spin lattice?

\section{Algorithm Formalism}

\begin{figure}
\centering
\includegraphics[width=3.4in]{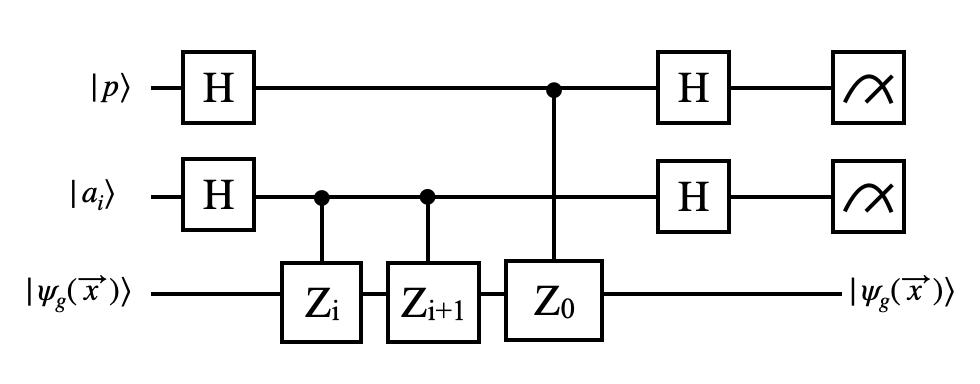}
\caption{Quantum circuit describing the Hadamard test to obtain $\langle\sigma_z^i\sigma_z^{i+1} \rangle$ on auziliary qubit $|a_i\rangle$ and parity of the first qubit $\langle \sigma_z^0\rangle$ on auxiliary qubit $|p\rangle$ from the ground state $|\psi_g(\vec{x})\rangle$.}
\label{fig:circuit}
\end{figure}

We obtain the Eigen spectra of the Hamiltonian for each data vector $\vec{x}$ and extract the ground state Eigen vector, $|\psi_g(\vec{x})\rangle$. The Eigen states of the Ising Hamiltonian are binary strings which reveal the low energy spin configuration on the lattice. We detect short-range spin order by evaluating the pairwise spin correlation function,
\begin{equation}
C = \frac{1}{N}\sum_i^N \langle \psi_g(\vec{x})|\sigma_{i}^z\sigma_{i+\Delta}^z|\psi_g(\vec{x})\rangle,
\end{equation}
where $\sigma^z_i$ is the Pauli z-operator for site $i$ and $\Delta$ are the neighbors of site $i$. A lattice with ferromagnetic order yields a spin correlation $C=1$, and a lattice with antiferromagnetic order yields a spin correlation $C=-1$. We find that the ground state of the feature map demonstrates an ordered configuration for both 1D and 2D lattice systems.

Projecting the feature space onto the Hamiltonian ground state provides a low dimensional representation of the data. The data sets are finite dimensional and we find the ground state is non-degenerate in all tested cases. There exist two possible antiferromagnetic states with opposite spin configurations 
\begin{align}
|\psi_g(\vec{x})\rangle &\rightarrow |AF1\rangle = |\uparrow\downarrow\uparrow\downarrow...\rangle\\
|\psi_g(\vec{x})\rangle &\rightarrow |AF2\rangle =|\downarrow\uparrow\downarrow\uparrow...\rangle.
\end{align}
Ground state information is extractable algorithmically on a quantum computer. Ising model ground states are prepared using quantum approximate optimization algorithm (QAOA)~\cite{FarhiQAOA}, imaginary time evolution~\cite{Motta2020}, or algorithmic cooling~\cite{Metcalf2022, Polla2021}. We use non-destructive measurements on the system qubits to extract information about the feature map ground state. The ground state correlation is extracted from the wave function using $N-1$ auxiliary qubits Fig~\ref{fig:circuit}. In addition, we propose an additional auxiliary qubit to check the parity on the first system qubit in order to distinguish between the two antiferromagnetic states. The antiferromagnetic state is identified by a correlation $C=-1$ and the two antiferromagnetic states are distinguished by the parity, 
\begin{align}
\langle \sigma_z^0\rangle &= 1 \rightarrow |AF1\rangle\\
\langle \sigma_z^0\rangle &= -1 \rightarrow |AF2\rangle.
\end{align}
We can compose a classical, joint probability $p(\psi_g(\vec{x}),y)$ spanning the possible ground state configurations and the class label $y$. There are six possible states normalized over a subset of labeled data $N_{train},$
\begin{align}
p(AF1,0) &= \frac{N_{AF1}^0}{N_{train}}\\
p(AF2,0) &= \frac{N_{AF2}^0}{N_{train}}\\
p(\mathcal{R},0) &= \frac{N_{\mathcal{R}}^0}{N_{train}}\\
p(AF1,1) &= \frac{N_{AF1}^1}{N_{train}}\\
p(AF2,1) &= \frac{N_{AF2}^1}{N_{train}}\\
p(\mathcal{R},1) &= \frac{N_{\mathcal{R}}^1}{N_{train}}
\end{align}
where $\mathcal{R}$ is the residual state space when $C\neq-1$.  The composite probability distribution is used to predict the label of an unclassified data vector $\vec{x}$ after determining the ground state solution in Ising feature space. Two possible states remain after tracing out the excluded states each associated with a separate label. The probabilities of these two states are used to predict the class of the unknown data vector.

The labeled data probabilities are separated and the distributions compared to understand the effectiveness of our approach on data sets. The total variation distance (TVD) is a distance metric to compare the distributions between classes,
\begin{equation}
TVD(P,Q) = \frac{1}{2}\sum_i^{N} |p_i - q_i|,
\end{equation}
where $p_i$ and $q_i$ are state probabilities. We use the TVD metric to quantify the distribution distance of the two data classes and use this information to predict the performance of the quantum classification tasks when data is encoded on varying spin lattices.

\section{Numerical Results}

We chose three publicly available data sets typical for testing binary classification models. The Kaggle hosted credit card transaction data set and the e-commerce data set are used to benchmark models at detecting fraud. Classical machine learning algorithms perform far better on the credit card transaction data set, while the e-commerce dataset is more difficult to classify. One intention of our work is to gain insights to why the e-commerce data set is more difficult by evaluating its expression on the spin lattice. In addition to the fraud data sets, we evaluate the UCI breast cancer data set to verify that our results are general to other forms of data. Data is down sampled from the full set to obtain a balanced subset. Larger samples of $N_{train}$ will reduce sampling error in the distributions with an overhead of more circuit simulations. We perform Principal Component Analysis (PCA) to select the principal components of the feature space. These principal components are a linear combination of the features resulting from an singular value decomposition of the data. The data is scaled between $(-a,a)$ where $a$ is the scaling factor. A data vector $\vec{x}$ is mapped to the Hamiltonian in Equation~\ref{Eq:Hamiltonian} with $S=\{1,2\}$ for three different lattice configurations: 1D chain, 2D square lattice, 2D triangle lattice. 

\begin{figure*}[t!]
\includegraphics[width=\textwidth]{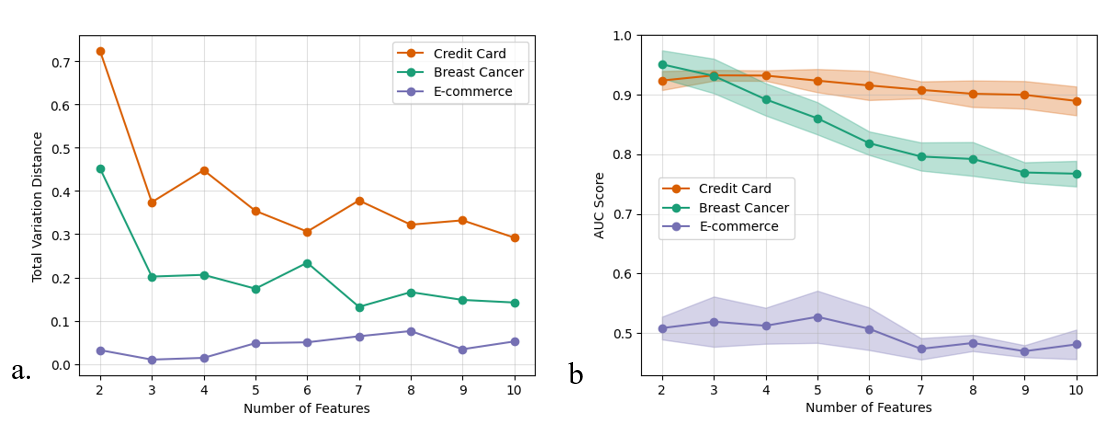}
\caption{a) Total variation distance between distributions of antiferromagnetic ground states associated with separate classes for increasing number of features.b) AUC score of quantum support vector machine for increasing feature number. This was calculated for 3 different sets of binary classification data, each of which were balanced then sampled down to 200 data points. Each set was encoded on one-dimensional lattice using linear entangling map.}
\label{fig:linear_map_tvd}
\end{figure*}


\begin{figure}
\centering
\includegraphics[width=3.4in]{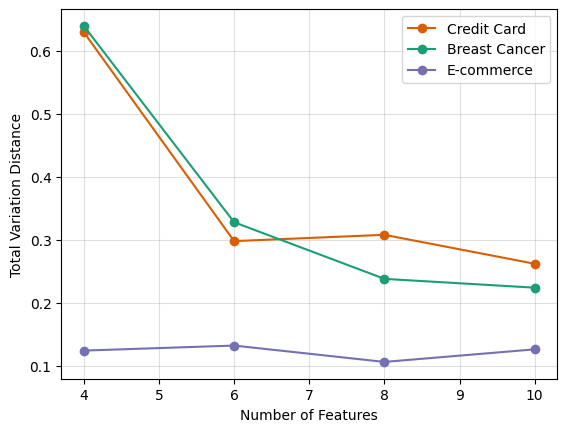}
\caption{Total variation distance between distributions of antiferromagnetic ground states associated with separate classes for increasing number of features. This was calculated for 3 different sets of binary classification data, each of which were balanced then sampled down to 200 data points. Each set was encoded on 2-dimensional square lattice ladder entangling map.}
\label{fig:square_map_tvd}
\end{figure}

\begin{figure*}[t!]
\centering
\includegraphics[width=\textwidth]{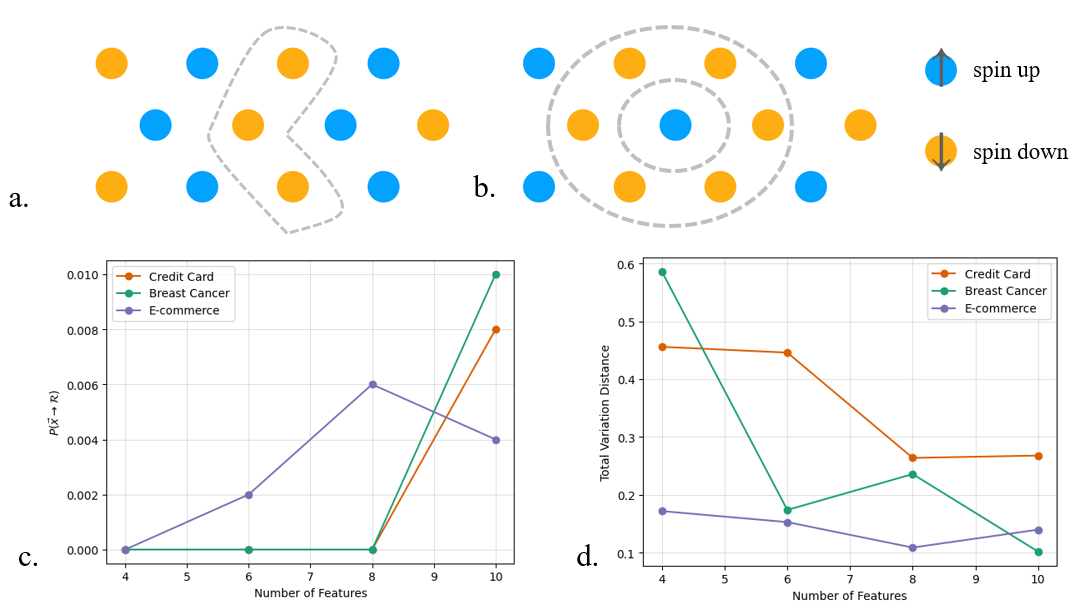}
\caption{a) Zig-zag ferromagnetic order b) graphene-like ferromagnetic order on $3\times4$ triangle lattice. Blue circles represent spin up and yellow circles represent spin down. c) Probability of residual states $P(\mathcal{R})$ for the three data sets encoded onto a triangle lattice. d) Total variation distance between distributions of antiferromagnetic ground states associated with separate classes for increasing number of features. This was calculated for 3 different sets of binary classification data, each of which were balanced then sampled down to 200 data points. Each set was encoded on triangle lattice ladder entangling map.}
\label{fig:triangle_map_Pr}
\end{figure*}



\textit{One-Dimensional Encoding --} The ground state of the linear entanglement map is \textit{always} anti-correlated such that $P(\vec{x}\rightarrow \mathcal{R}) = 0$. We evaluate the TVD between distributions associated with separate classes (e.g. fraudulent vs. genuine and malignant vs. benign) in Figure~\ref{fig:linear_map_tvd}(a). TVD is largest for the credit card transaction data set, meaning the distribution of the fraudulent transactions to the antiferromagnetic state space is considerably different than the distribution of genuine transactions. Thus, the Ising model is capable of discriminating the two classes, albeit with low probability. In contrast, the distribution difference for the e-commerce data is near zero and the Ising model is unable to discriminate the classes. This is interesting because the e-commerce data is known to challenge classical models as well. Our model is a technique to demonstrate the similarity between the classes in high-dimensional non-linear space. One expects the data non-linearity is very complex and higher-order non-linear models are needed to discriminate the data. The breast cancer data set falls in between the TVDs of the two other data sets, so the Ising model is still able to discriminate between malignant and benign instances. We expect, however, the quantum model to perform worse on this data set compared the credit card fraud data because the TVD is reduced. 

We perform quantum support vector classification tasks on the three data sets to extrapolate meaning behind the linear entangling map TVD. We hypothesize better model discrimination when the data set exhibits a large TVD. The Qiskit and Scikit-learn libraries are used to build the quantum and classical machine learning algorithms~\cite{Qiskit,sklrn}. The objective is not to find the best fit model, rather to determine the QML algorithm performance given our knowledge of the distribution difference.

Quantum kernels evaluate the distance between data vectors by determining the overlap between the data points in feature space
\begin{equation}
K_{ij} = \langle \phi(\vec{x}_j)|\phi(\vec{x}_i)\rangle. 
\end{equation}
The kernel, obtained by rotating in the Eigen basis of the encoded spin lattice, inputs to a classical Support Vector Classifer (SVC) which optimizes the support vectors to separate the training data classes. We find support vector classifiers have superior performance to heuristic quantum neural networks in classification tasks. Area under the curve (AUC) determines the model discrimination quality to correctly classify new data vectors. We evaluate and compare the AUC metric using an SVC with a quantum kernel that evaluates data distance in the Ising lattice Eigen space. The quality of this metric is correlated to the ground state distribution TVD.


We sample the data, undersampling the majority class, to obtain a balanced subset and partition the sample into training and test sets. We leave hyper-parameters fixed for all three data sets. The scaling parameter $a = 1$ for all data sets. We evaluate the kernel performance for increasing number of principle component vectors mapped to qubits using the linear chain Ising encoding Fig~\ref{fig:linear_map_tvd}(b). The QSVC performs better on the credit card transaction data than the breast cancer dataset which is contrary to the performance of the classical SVC but inline with the ground state distribution TVD. The QSVC performs worst on the E-commerce dataset. Our results, imply the ground state of the feature map can provide insight into the outcome of quantum classification algorithms using the Ising type feature map. 


\textit{Two-Dimensional Encoding --} The resulting anti-correlated ground state extends to data encoded on two dimesnional spin lattices. Ground states of a data vector mapped to a square ladder results in perfectly anti-correlated spin states on the ladder. TVDs of the breast cancer data and credit card fraud data are similar, and we see a reduced TVD from the one-dimensional chain encoding Fig~\ref{fig:square_map_tvd}. Preference for nearest neighbor entanglement on a chain could result from the interplay of principal component analysis and the interplay of non-linear feature expression. As a result, the QSVM has worse performance using the square ladder entanglement strategy.

There are cases where $P(x\rightarrow\mathcal{R}) > 0$ and the ground states are no longer anti-correlated. Such is the case with the triangle lattice entangling map. Antiferromagnetic ground states on triangle lattices exhibit geometric frustration characterized by high ground state degeneracy\cite{Ramirez2003}. The disorder from the classical data encoded on the triangle lattice breaks this degeneracy, and we find one of the ordered antiferromagnetic states as a solution to the data encoding. Evaluating the probability of the residual state space on the triangle ladder show a size dependence for both the breast cancer data and the credit card fraud data Fig~\ref{fig:triangle_map_Pr}(c). The ecommerce data, in contrast, exhibits residual probability for small lattice sizes. For all of these data sets $P(\mathcal{R})$ is small. We extend our model into a fully 2D model and characterize the spin order on a $3\times4$ triangle lattice. The most typical state is the zig-zag phase on the triangle lattice Fig~\ref{fig:triangle_map_Pr}(a). In addition, we find certain instances have a hexagonal, graphene-like ferromagnetic order on the spin lattice Fig~\ref{fig:triangle_map_Pr}(b). This means our data vectors subside in different sectors of a ground state phase diagram with different antiferromagnetic ordering. Connecting the physics of these phases with the classical data vectors that generate them can potentially lead to new insights about the data. The TVD for the triangle lattice diminishes rapidly for the datasets rather than stabilizing like the square lattice and one-dimensional chain Fig~\ref{fig:triangle_map_Pr}(d).

\section{Order Stability and Generalization}

Stability of the antiferromagnetic state results in better performance of the ZZ feature map. Our original curiosity arose due to the scaling parameter introduced in the time-evolution operator to improve quantum distance measures between data points (\cite{Park2020}), and we sought to determine the physical meaning behind this scaling parameter.

\begin{figure}[h!]
\centering
\includegraphics[width=3.4in]{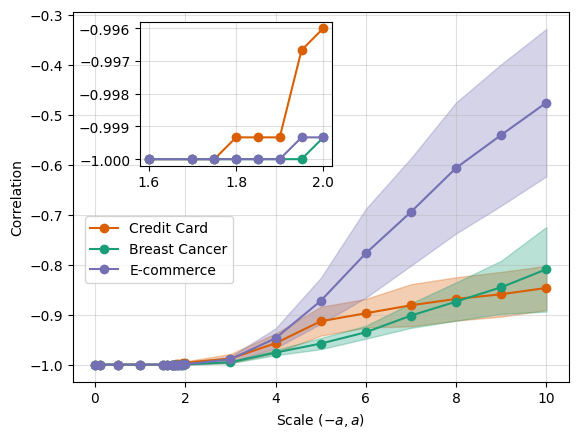}
\caption{One-dimensional linear entangling map ground state spin correlation with increasing scaling parameter $a$. Shown for 3 different datasets, each of which were balanced then sampled down to 200 data points and 4 features. (Inset) higher resilution plot to examine order-disorder transition point $a_0$.}
\label{fig:corr_scale_chain}
\end{figure}

\textit{Parameter Scaling -- }Quality of the Ising feature map reduces significantly if data isn't scaled appropriately. Ref~\cite{Park2020} introduces a scaling hyper-parameter $a$ in the unitary evolution operator,
\begin{equation}
U(a, \vec{x}) = e^{iaH(\vec{x})}.
\end{equation}
We evaluate the spin correlation of the ground state of scaled data encoded on the 1-dimensional linear entangling map Fig~\ref{fig:corr_scale_chain}. The antiferromagnetic order collapses as data is scaled to larger values seen by the diminishing correlation. The credit card fraud data de-correlated at a lesser value $a$ than the breast cancer and e-commerce data. In particular, the state space distributions of the two data classes of data differ enough that classification is no longer possible. The degree of divergence beyond this point is data dependent. The e-commerce dataset is less robust to the scaling, therefore, it diverges at a much faster rate in comparison to both the credit card and breast cancer datasets. 

The destruction of correlation in the model is likely arises from strong lattice disorder, thus, the lattice potential no longer favors antiferromagnetic order. This transition point, $a_{0}$, appears inconsistent between datasets though very similar, perhaps due to a physical phase transition. Although initial thoughts assumed this transition would happen around $a_{0}=\pi$, it is clear this instability occurs at a smaller scaling parameter. The physical meaning behind this would be an interesting topic for further research.

\begin{figure}[h!]
\centering
\includegraphics[width=3.4in]{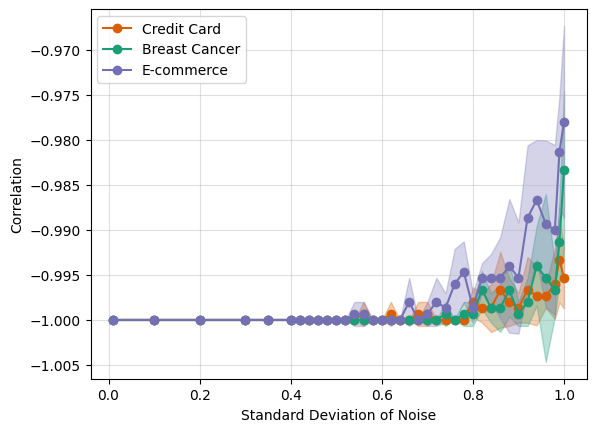}
\caption{Spin correlation with increasing standard deviation of Gaussian noise added to the data.}
\label{fig:corr_noise_chain}
\end{figure}

\textit{Noise --} The role of noise in data is important for model generalization. We assess the stability of antiferromagnetic order when Gaussian noise is added to the data Fig~\ref{fig:corr_noise_chain}. We see that the spin correlation of the ground states begins to diverge for higher levels of noise. The antiferromagnetic nature of the ground state is robust to Gaussian noise with a standard deviation $\sigma \leq 0.5$, however, this is broken for highly noisy data. Similarly to scaling, we see the E-commerce data set is less robust to noise.

\section{Conclusion}
Non-degenerate antiferromagnetic ground states result from the disordered Ising Hamiltonian encoded with classical data. Binary classes are dicriminatable by evaluating the distribution of training samples on the state space; this means we can predict the resulting ground state of a classical data vector with some degree of precision. We propose a quantum sub-protocol to extract ground state information to build the statistical ensemble. Our results are general to both one dimensional and two dimensional Ising lattices. Further meaning is gained by relating the distribution difference between two classes and the quality of quantum kernel methods, and we find a dataset with a higher TVD translates to improved model discrimination. An interesting direction for future research is to optimize the TVD to fully separate the distributions in the antiferromagnetic state space. This will likely improve the feature encoding for other machine learning primitives as well and allow for a more accurate predictive framework.

Anti-correlated behavior across the chain is diminished due to geometric frustration in the triangle lattice leading to a higher probability of residual state space. Further the anti-ferromagnetic ground state stability is dependent on data scaling which provides physical meaning behind the need for a scaling hyper-parameter in Ising Hamiltonian encoding. The role of disorder in ground state stability leaves room for further investigations on the role of entropy in data encodings and stability of correlated states. We also find the antiferromagnetic state is robust to Gaussian noise with standard deviation $\sigma \leq 0.5$. In the case of the credit card transaction data, it outperformed the classical SVC even with a small number of features.

Mapping classical data to Ising Hamiltonians results in unique, correlated states. These resulting correlated states provide physical intuition behind classical data in machine learning algorithms. In fact, the physics can inform whether the quantum machine learning algorithm will perform well. Our results explain why the Ising encoding works well for certain data sets and not for others. It also explains the encoding robustness to data engineering techniques like scaling and robustness to data noise. Our results enable a more interpretable framework for encoding classical data on quantum computers by examining the underlying spin physics. The evaluation of classical data encoded to quantum Hamiltonians like the Heisenberg model, will likely result in further unique physics features arising from lattice disorder. We leave further model evaluation for future work.

\textit{DISLCAIMER: }
This paper was prepared for information purposes, and is not a product of HSBC Europe or its affiliates. Neither HSBC Europe nor any of its affiliates make any explicit or implied representation or warranty and none of them accept any liability in connection with this paper, including, but not limited to, the completeness, accuracy, reliability of information contained herein and the potential legal, compliance, tax or accounting effects thereof.  Copyright HSBC Group 2023.

\bibliographystyle{apsrev4-1}
\bibliography{reference2}

\begin{thebibliography}{21}%
\makeatletter
\providecommand \@ifxundefined [1]{%
 \@ifx{#1\undefined}
}%
\providecommand \@ifnum [1]{%
 \ifnum #1\expandafter \@firstoftwo
 \else \expandafter \@secondoftwo
 \fi
}%
\providecommand \@ifx [1]{%
 \ifx #1\expandafter \@firstoftwo
 \else \expandafter \@secondoftwo
 \fi
}%
\providecommand \natexlab [1]{#1}%
\providecommand \enquote  [1]{``#1''}%
\providecommand \bibnamefont  [1]{#1}%
\providecommand \bibfnamefont [1]{#1}%
\providecommand \citenamefont [1]{#1}%
\providecommand \href@noop [0]{\@secondoftwo}%
\providecommand \href [0]{\begingroup \@sanitize@url \@href}%
\providecommand \@href[1]{\@@startlink{#1}\@@href}%
\providecommand \@@href[1]{\endgroup#1\@@endlink}%
\providecommand \@sanitize@url [0]{\catcode `\\12\catcode `\$12\catcode
  `\&12\catcode `\#12\catcode `\^12\catcode `\_12\catcode `\%12\relax}%
\providecommand \@@startlink[1]{}%
\providecommand \@@endlink[0]{}%
\providecommand \url  [0]{\begingroup\@sanitize@url \@url }%
\providecommand \@url [1]{\endgroup\@href {#1}{\urlprefix }}%
\providecommand \urlprefix  [0]{URL }%
\providecommand \Eprint [0]{\href }%
\providecommand \doibase [0]{http://dx.doi.org/}%
\providecommand \selectlanguage [0]{\@gobble}%
\providecommand \bibinfo  [0]{\@secondoftwo}%
\providecommand \bibfield  [0]{\@secondoftwo}%
\providecommand \translation [1]{[#1]}%
\providecommand \BibitemOpen [0]{}%
\providecommand \bibitemStop [0]{}%
\providecommand \bibitemNoStop [0]{.\EOS\space}%
\providecommand \EOS [0]{\spacefactor3000\relax}%
\providecommand \BibitemShut  [1]{\csname bibitem#1\endcsname}%
\let\auto@bib@innerbib\@empty
\bibitem [{\citenamefont {Schuld}\ \emph {et~al.}(2021)\citenamefont {Schuld},
  \citenamefont {Sweke},\ and\ \citenamefont {Meyer}}]{Schuld2021}%
  \BibitemOpen
  \bibfield  {author} {\bibinfo {author} {\bibfnamefont {M.}~\bibnamefont
  {Schuld}}, \bibinfo {author} {\bibfnamefont {R.}~\bibnamefont {Sweke}}, \
  and\ \bibinfo {author} {\bibfnamefont {J.~J.}\ \bibnamefont {Meyer}},\
  }\href@noop {} {\bibfield  {journal} {\bibinfo  {journal} {Phys. Rev. A}\
  }\textbf {\bibinfo {volume} {103}},\ \bibinfo {pages} {032430} (\bibinfo
  {year} {2021})}\BibitemShut {NoStop}%
\bibitem [{\citenamefont {Peters}\ and\ \citenamefont
  {Schuld}(2022)}]{Peters2022}%
  \BibitemOpen
  \bibfield  {author} {\bibinfo {author} {\bibfnamefont {E.}~\bibnamefont
  {Peters}}\ and\ \bibinfo {author} {\bibfnamefont {M.}~\bibnamefont
  {Schuld}},\ }\href@noop {} {\bibfield  {journal} {\bibinfo  {journal}
  {ArXiv}\ ,\ \bibinfo {pages} {2209.05523}} (\bibinfo {year}
  {2022})}\BibitemShut {NoStop}%
\bibitem [{\citenamefont {Schuld}\ and\ \citenamefont
  {Petruccione}(2018)}]{SchuldPetruccioneBook}%
  \BibitemOpen
  \bibfield  {author} {\bibinfo {author} {\bibfnamefont {M.}~\bibnamefont
  {Schuld}}\ and\ \bibinfo {author} {\bibfnamefont {F.}~\bibnamefont
  {Petruccione}},\ }\href@noop {} {\emph {\bibinfo {title} {Supervised Learning
  with Quantum Computers}}}\ (\bibinfo  {publisher} {Springer},\ \bibinfo
  {year} {2018})\BibitemShut {NoStop}%
\bibitem [{\citenamefont {Havlicek}\ \emph {et~al.}(2019)\citenamefont
  {Havlicek}, \citenamefont {Córcoles}, \citenamefont {Temme}, \citenamefont
  {Harrow}, \citenamefont {Kandala}, \citenamefont {Chow},\ and\ \citenamefont
  {Gambetta}}]{Havlicek2019}%
  \BibitemOpen
  \bibfield  {author} {\bibinfo {author} {\bibfnamefont {V.}~\bibnamefont
  {Havlicek}}, \bibinfo {author} {\bibfnamefont {A.~D.}\ \bibnamefont
  {Córcoles}}, \bibinfo {author} {\bibfnamefont {K.}~\bibnamefont {Temme}},
  \bibinfo {author} {\bibfnamefont {A.~W.}\ \bibnamefont {Harrow}}, \bibinfo
  {author} {\bibfnamefont {A.}~\bibnamefont {Kandala}}, \bibinfo {author}
  {\bibfnamefont {J.~M.}\ \bibnamefont {Chow}}, \ and\ \bibinfo {author}
  {\bibfnamefont {J.~M.}\ \bibnamefont {Gambetta}},\ }\href@noop {} {\bibfield
  {journal} {\bibinfo  {journal} {Nature}\ }\textbf {\bibinfo {volume} {567}}
  (\bibinfo {year} {2019})}\BibitemShut {NoStop}%
\bibitem [{\citenamefont {Gilpin}\ \emph {et~al.}(2018)\citenamefont {Gilpin},
  \citenamefont {Bau}, \citenamefont {Yuan}, \citenamefont {Bajwa},
  \citenamefont {Specter},\ and\ \citenamefont {Kagal}}]{Gilpin2018}%
  \BibitemOpen
  \bibfield  {author} {\bibinfo {author} {\bibfnamefont {L.~H.}\ \bibnamefont
  {Gilpin}}, \bibinfo {author} {\bibfnamefont {D.}~\bibnamefont {Bau}},
  \bibinfo {author} {\bibfnamefont {B.~Z.}\ \bibnamefont {Yuan}}, \bibinfo
  {author} {\bibfnamefont {A.}~\bibnamefont {Bajwa}}, \bibinfo {author}
  {\bibfnamefont {M.}~\bibnamefont {Specter}}, \ and\ \bibinfo {author}
  {\bibfnamefont {L.}~\bibnamefont {Kagal}},\ }\href@noop {} {\bibfield
  {journal} {\bibinfo  {journal} {IEEE 5th International Conference on data
  science and advanced analytics (DSAA)}\ } (\bibinfo {year}
  {2018})}\BibitemShut {NoStop}%
\bibitem [{\citenamefont {Rudin}(2019)}]{Rudin2019}%
  \BibitemOpen
  \bibfield  {author} {\bibinfo {author} {\bibfnamefont {C.}~\bibnamefont
  {Rudin}},\ }\href@noop {} {\bibfield  {journal} {\bibinfo  {journal} {Nature
  Machine Intelligence}\ }\textbf {\bibinfo {volume} {1}},\ \bibinfo {pages}
  {206–215} (\bibinfo {year} {2019})}\BibitemShut {NoStop}%
\bibitem [{UCI()}]{UCIBCData}%
  \BibitemOpen
  \href@noop {} {\enquote {\bibinfo {title} {Uci machine learning repository:
  Breast cancer data set},}\ }\bibinfo {howpublished}
  {\url{https://archive.ics.uci.edu/ml/datasets/Breast20Cancer}}\BibitemShut
  {NoStop}%
\bibitem [{Kag()}]{KaggleCCT}%
  \BibitemOpen
  \href@noop {} {\enquote {\bibinfo {title} {Kaggle credit card fraud
  detection},}\ }\bibinfo {howpublished}
  {\url{https://www.kaggle.com/datasets/mlg-ulb/creditcardfraud}}\BibitemShut
  {NoStop}%
\bibitem [{Eco()}]{EcommData}%
  \BibitemOpen
  \href@noop {} {\enquote {\bibinfo {title} {Kaggle fraud e-commerce},}\
  }\bibinfo {howpublished}
  {\url{https://www.kaggle.com/vbinh002/fraud-ecommerce}}\BibitemShut {NoStop}%
\bibitem [{\citenamefont {Schuld}\ and\ \citenamefont
  {Killoran}(2019)}]{Schuld2019}%
  \BibitemOpen
  \bibfield  {author} {\bibinfo {author} {\bibfnamefont {M.}~\bibnamefont
  {Schuld}}\ and\ \bibinfo {author} {\bibfnamefont {N.}~\bibnamefont
  {Killoran}},\ }\href@noop {} {\bibfield  {journal} {\bibinfo  {journal}
  {Phys. Rev. Lett.}\ }\textbf {\bibinfo {volume} {122}} (\bibinfo {year}
  {2019})}\BibitemShut {NoStop}%
\bibitem [{\citenamefont {Altares-Lopez}\ \emph {et~al.}(2021)\citenamefont
  {Altares-Lopez}, \citenamefont {Ribeiro},\ and\ \citenamefont
  {Garcia-Ripoll}}]{AL2021}%
  \BibitemOpen
  \bibfield  {author} {\bibinfo {author} {\bibfnamefont {S.}~\bibnamefont
  {Altares-Lopez}}, \bibinfo {author} {\bibfnamefont {A.}~\bibnamefont
  {Ribeiro}}, \ and\ \bibinfo {author} {\bibfnamefont {J.~J.}\ \bibnamefont
  {Garcia-Ripoll}},\ }\href@noop {} {\bibfield  {journal} {\bibinfo  {journal}
  {Quantum Science and Technology}\ }\textbf {\bibinfo {volume} {6}},\ \bibinfo
  {pages} {045015} (\bibinfo {year} {2021})}\BibitemShut {NoStop}%
\bibitem [{\citenamefont {Massimiliano~Incudini}(2022)}]{Incudini2022}%
  \BibitemOpen
  \bibfield  {author} {\bibinfo {author} {\bibfnamefont {A.~D.~P.}\
  \bibnamefont {Massimiliano~Incudini}, \bibfnamefont {Francesco~Martini}},\
  }\href@noop {} {\bibfield  {journal} {\bibinfo  {journal} {ArXiv}\ ,\
  \bibinfo {pages} {2209.11144}} (\bibinfo {year} {2022})}\BibitemShut
  {NoStop}%
\bibitem [{\citenamefont {Torabian}\ and\ \citenamefont
  {Krems}(2022)}]{Torabian2022}%
  \BibitemOpen
  \bibfield  {author} {\bibinfo {author} {\bibfnamefont {E.}~\bibnamefont
  {Torabian}}\ and\ \bibinfo {author} {\bibfnamefont {R.~V.}\ \bibnamefont
  {Krems}},\ }\href@noop {} {\bibfield  {journal} {\bibinfo  {journal} {ArXiv}\
  ,\ \bibinfo {pages} {2203.13848}} (\bibinfo {year} {2022})}\BibitemShut
  {NoStop}%
\bibitem [{\citenamefont {Farhi}\ \emph {et~al.}(2014)\citenamefont {Farhi},
  \citenamefont {Goldstone},\ and\ \citenamefont {Gutmann}}]{FarhiQAOA}%
  \BibitemOpen
  \bibfield  {author} {\bibinfo {author} {\bibfnamefont {E.}~\bibnamefont
  {Farhi}}, \bibinfo {author} {\bibfnamefont {J.}~\bibnamefont {Goldstone}}, \
  and\ \bibinfo {author} {\bibfnamefont {S.}~\bibnamefont {Gutmann}},\
  }\href@noop {} {\bibfield  {journal} {\bibinfo  {journal} {ArXiv}\ ,\
  \bibinfo {pages} {1411.4028}} (\bibinfo {year} {2014})}\BibitemShut {NoStop}%
\bibitem [{\citenamefont {Motta}\ \emph {et~al.}(2020)\citenamefont {Motta},
  \citenamefont {Sun}, \citenamefont {Tan}, \citenamefont {Rourke},
  \citenamefont {Ye}, \citenamefont {Minnich}, \citenamefont {Brandao},\ and\
  \citenamefont {Chan}}]{Motta2020}%
  \BibitemOpen
  \bibfield  {author} {\bibinfo {author} {\bibfnamefont {M.}~\bibnamefont
  {Motta}}, \bibinfo {author} {\bibfnamefont {C.}~\bibnamefont {Sun}}, \bibinfo
  {author} {\bibfnamefont {A.~T.~K.}\ \bibnamefont {Tan}}, \bibinfo {author}
  {\bibfnamefont {M.~J.~O.}\ \bibnamefont {Rourke}}, \bibinfo {author}
  {\bibfnamefont {E.}~\bibnamefont {Ye}}, \bibinfo {author} {\bibfnamefont
  {A.~J.}\ \bibnamefont {Minnich}}, \bibinfo {author} {\bibfnamefont {F.~G.
  S.~L.}\ \bibnamefont {Brandao}}, \ and\ \bibinfo {author} {\bibfnamefont
  {G.~K.-L.}\ \bibnamefont {Chan}},\ }\href@noop {} {\bibfield  {journal}
  {\bibinfo  {journal} {Nature Physics}\ }\textbf {\bibinfo {volume} {16}},\
  \bibinfo {pages} {205} (\bibinfo {year} {2020})}\BibitemShut {NoStop}%
\bibitem [{\citenamefont {Metcalf}\ \emph {et~al.}(2022)\citenamefont
  {Metcalf}, \citenamefont {Stone}, \citenamefont {Klymko}, \citenamefont
  {Kemper}, \citenamefont {Sarovar},\ and\ \citenamefont
  {de~Jong}}]{Metcalf2022}%
  \BibitemOpen
  \bibfield  {author} {\bibinfo {author} {\bibfnamefont {M.}~\bibnamefont
  {Metcalf}}, \bibinfo {author} {\bibfnamefont {E.}~\bibnamefont {Stone}},
  \bibinfo {author} {\bibfnamefont {K.}~\bibnamefont {Klymko}}, \bibinfo
  {author} {\bibfnamefont {A.~F.}\ \bibnamefont {Kemper}}, \bibinfo {author}
  {\bibfnamefont {M.}~\bibnamefont {Sarovar}}, \ and\ \bibinfo {author}
  {\bibfnamefont {W.~A.}\ \bibnamefont {de~Jong}},\ }\href@noop {} {\bibfield
  {journal} {\bibinfo  {journal} {Quantum Science and Technology}\ }\textbf
  {\bibinfo {volume} {7}},\ \bibinfo {pages} {025017} (\bibinfo {year}
  {2022})}\BibitemShut {NoStop}%
\bibitem [{\citenamefont {Polla}\ \emph {et~al.}(2021)\citenamefont {Polla},
  \citenamefont {Herasymenko},\ and\ \citenamefont {O'Brien}}]{Polla2021}%
  \BibitemOpen
  \bibfield  {author} {\bibinfo {author} {\bibfnamefont {S.}~\bibnamefont
  {Polla}}, \bibinfo {author} {\bibfnamefont {Y.}~\bibnamefont {Herasymenko}},
  \ and\ \bibinfo {author} {\bibfnamefont {T.~E.}\ \bibnamefont {O'Brien}},\
  }\href@noop {} {\bibfield  {journal} {\bibinfo  {journal} {Phys. Rev. A}\
  }\textbf {\bibinfo {volume} {104}},\ \bibinfo {pages} {012414} (\bibinfo
  {year} {2021})}\BibitemShut {NoStop}%
\bibitem [{Qis()}]{Qiskit}%
  \BibitemOpen
  \href@noop {} {\enquote {\bibinfo {title} {Qiskit},}\ }\bibinfo
  {howpublished} {\url{https://www.qiskit.org}}\BibitemShut {NoStop}%
\bibitem [{skl()}]{sklrn}%
  \BibitemOpen
  \href@noop {} {\enquote {\bibinfo {title} {Scikit learn},}\ }\bibinfo
  {howpublished} {\url{https://www.scikit-learn.org}}\BibitemShut {NoStop}%
\bibitem [{\citenamefont {Ramirez}(2003)}]{Ramirez2003}%
  \BibitemOpen
  \bibfield  {author} {\bibinfo {author} {\bibfnamefont {A.}~\bibnamefont
  {Ramirez}},\ }\href@noop {} {\bibfield  {journal} {\bibinfo  {journal}
  {Nature}\ }\textbf {\bibinfo {volume} {421}},\ \bibinfo {pages} {483}
  (\bibinfo {year} {2003})}\BibitemShut {NoStop}%
\bibitem [{\citenamefont {Park}\ \emph {et~al.}(2020)\citenamefont {Park},
  \citenamefont {Quanz}, \citenamefont {Wood}, \citenamefont {Higgins},\ and\
  \citenamefont {Harishankar}}]{Park2020}%
  \BibitemOpen
  \bibfield  {author} {\bibinfo {author} {\bibfnamefont {J.-E.}\ \bibnamefont
  {Park}}, \bibinfo {author} {\bibfnamefont {B.}~\bibnamefont {Quanz}},
  \bibinfo {author} {\bibfnamefont {S.}~\bibnamefont {Wood}}, \bibinfo {author}
  {\bibfnamefont {H.}~\bibnamefont {Higgins}}, \ and\ \bibinfo {author}
  {\bibfnamefont {R.}~\bibnamefont {Harishankar}},\ }\href@noop {} {\bibfield
  {journal} {\bibinfo  {journal} {34th Conference on Neural Information
  Processing Systems}\ ,\ \bibinfo {pages} {First Workshop on Quantum Tensor
  Networks in Machine Learning}} (\bibinfo {year} {2020})}\BibitemShut
  {NoStop}%
\end{thebibliography}%
\end{document}